\documentclass[a4paper,11pt]{article}
\pdfoutput=1 
\usepackage{jinstpub} 

\usepackage{float}
\usepackage{array}

\usepackage{etoolbox}
\usepackage{setspace}
\usepackage[colorlinks,citecolor=red,linkcolor=blue]{hyperref}

\newcounter{rowcntr}[table]
\renewcommand{\therowcntr}{\thetable.\arabic{rowcntr}}

\newcolumntype{N}{>{\refstepcounter{rowcntr}\therowcntr}c}

\AtBeginEnvironment{tabular}{\setcounter{rowcntr}{0}}

\title{Determination of charge spread, position resolution, energy resolution and gain uniformity of  Gas Electron Multipliers (GEM)}

\author[a,1]{Vishal Kumar,\note{Corresponding author.}}
\author[b]{Subhendu Das,}
\author[b]{Promita Roy,}
\author[c]{Purba Bhattacharyab,}
\author[b]{Supratik Mukhopadhyay,}
\author[b]{Nayana Majumdar,}
\author[b]{and Sandip Sarkar,\note{Retired Professor.}}

\affiliation[a]{Centre for Cosmology, Particle Physics and Phenomenology (CP3), Université catholique de Louvain, B-1348
Louvain la Neuve, Belgium}
\affiliation[b]{Saha Institute of Nuclear Physics, Kolkata - 700064, a CI institute of Homi Bhabha National Institute, Mumbai-400094, India}
\affiliation[c]{School of Basic and Applied Sciences, Adamas University, Kolkata-700126, India} 

\emailAdd{vishalkmrjswl@gmail.com}

\abstract{Gas electron multipliers (GEM) detectors are gaseous detectors widely used for tracking and imaging applications due to their good position resolution, high efficiency at high irradiation rates, among other factors. In the present work, position resolution, charge spread, energy resolution and gain uniformity have been investigated experimentally for single and double GEM geometries using an $^{55}$Fe source.
 The position resolution measurements have been performed by a novel method, using a high precision instrument for source movement and is found to be highly successful. The result shows that the double GEM can resolve positions with sigma values up to 36.8 $\mu m$ and 54.6 $\mu m$ in x and y directions, respectively.  To validate the experimental results, a Garfield simulation work has been carried out on charge spread.
}


\begin{document}
\maketitle
\flushbottom

\section{Introduction}
\label{S1}
Since the development of GEM detectors~\cite{Sauli:1997qp, Sauli:2016eeu}, high gain, good spatial resolution with high rate handling capacity and discharge free operation have made them very popular for particle tracking applications. Detectors base on GEMs have been installed in many well known experiments like in CMS endcap~\cite{Colaleo:2015vsq, Abbaneo:2014lja, malhotra2022gem} forward muon region, ALICE TPC~\cite{Lippmann:2014lay, lippmann2016continuous} replacing MWPC, COMPASS experiment~\cite{Altunbas:2002ds}, etc. Requirement of large area coverage in the upgrade of these major particle physics experiments has also been a pivotal reason for choosing GEM detectors.
Uniformity over the complete surface of the detector in terms of gain, position resolution and efficiency is crucial for the optimum performance of these detectors. In the case of particle tracking and imaging applications~\cite{bachmann2002high, Ahmed2022the, ju2016design}, it is important to obtain data that allows a meticulous reconstruction of the incident particle’s position and trajectory. Thus having good position resolution is one of the crucial factors to look for in proposed $/$ employed detectors.

Due to several advantages of GEM detectors, they are being employed in TPC applications~\cite{carnegie2005resolution, pinci2019high, adolfsson2021upgrade, fenker2008bonus}. Along with relatively good position resolution and high gain, it also helps in the reduction of ion back flow when used in stack~\cite{peskov2013ibf, bohmer2013simulation, lippmann2016continuous} or in a combination of other detectors like with 
Micromegas~\cite{dehmelt2017sphenix}, wire grid~\cite{wang2019studies} and bulk mesh~\cite{glaenzer2023reducing}. Their durability, high rate handling capability and sustained performance in high radiation environments has made them a detector of choice in system upgrades of CMS for the high-luminosity LHC. GEM detectors have been installed in CMS endcaps as  GE$1/1$~\cite{abbas2022quality, ivone2022commissioning, petre2022commissioning}, they are also in production, and going through performance and quality control tests for GE$2/1$~\cite{kim2022production, malhotra2022gem, pellecchia2023performance, cameron2022integration} upgrade and ME$0$~\cite{bianco2022high}. 
GEM detectors are proposed to be used in Soft X-Ray (SXR) tomography system for International Thermonuclear Experimental Reactor (ITER) oriented tokamaks to improve its existing performance~\cite{chernyshova2017development}. A successful trial has been conducted by GEM detectors at Tungsten (W) Environment in Steady-state Tokamak (WEST) for monitoring of tungsten impurities in plasma~\cite{mazon2022first}. 

The standard GEM foils used in GEM detectors are usually made of polyimide sheets 50~$\mu m$ in thickness with a 5~$\mu m$ copper cladding on each side~\cite{Sauli:2016eeu}. These foils have bi-conical holes with inner and outer diameters of 50 and 70~$\mu m$ respectively, and a center-to-center pitch of 140~$\mu m$ etched out in the copper-polyimide-copper sandwiched sheets in a hexagonal pattern.
The presence of polyimide and bi-conical hole structure aid GEM foil's optical transparency, gain and discharge-free operation at very high electric fields. 
These few tens of micron order holes act as electron multipliers which ensure low charge spread with high gain, thereby improving position resolution~\cite{Sauli:2016eeu}.

In the present work, a detailed investigation has been conducted to propose novel ways to measure / estimate charge spread, position resolution, energy resolution and gain uniformity. The experiments have been performed using existing resources in the laboratory rather than the conventional methods that require beam lines or x-ray generators for position resolution and gain uniformity measurements, respectively.
The following sections include these measurements along with comparison of our approach with the conventional methods and discussions of their benefits and drawbacks. The charge spread on the readout has been further verified by numerical studies performed using the 
Garfield simulations~\cite{Veenhof:1998tt} (involving Heed~\cite{Smirnov:2005yi}, Magboltz~\cite{Biagi:1999nwa} and neBEM~\cite{Majumdar:2006jf,muhkopadhyay2006computation}).



\section{Experimental setup}
\label{S2}
\begin{figure}[ht]
	\centering
	\includegraphics[width=7cm,keepaspectratio]{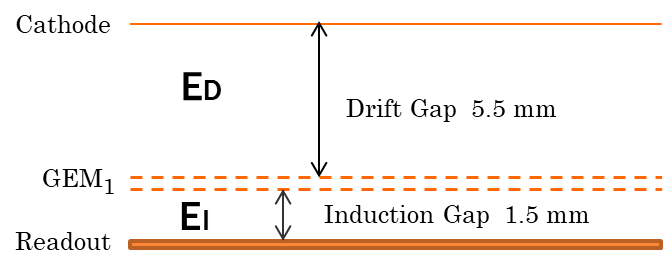} ~~
	\includegraphics[width=7cm,keepaspectratio]{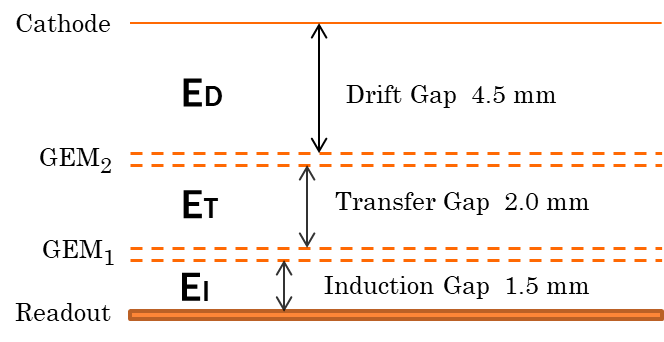}
	\caption{Schematic diagram of single (left) and double (right) GEM geometries.}
	\label{GEM_geometry}
\end{figure}

The experiments have been conducted on a 10 by 10 $cm$ standard GEM foil procured from CERN for both single and double GEM configuration. The detector assembly consists of GEM foil(s) placed in between a cathode plane and a readout anode. The geometrical configuration of the detector is as shown in figure~\ref{GEM_geometry}, the drift, transfer and induction gaps have been maintained with the help of spacers carved out of 500 $\mu m$ thick PCB material. These GEM foil(s) and cathode plane stacks are placed on a readout PCB with a gas-filled enclosure having a mylar sheet as top cover.

\begin{figure}[ht]
	\centering
	\includegraphics[height=4.5cm,keepaspectratio]{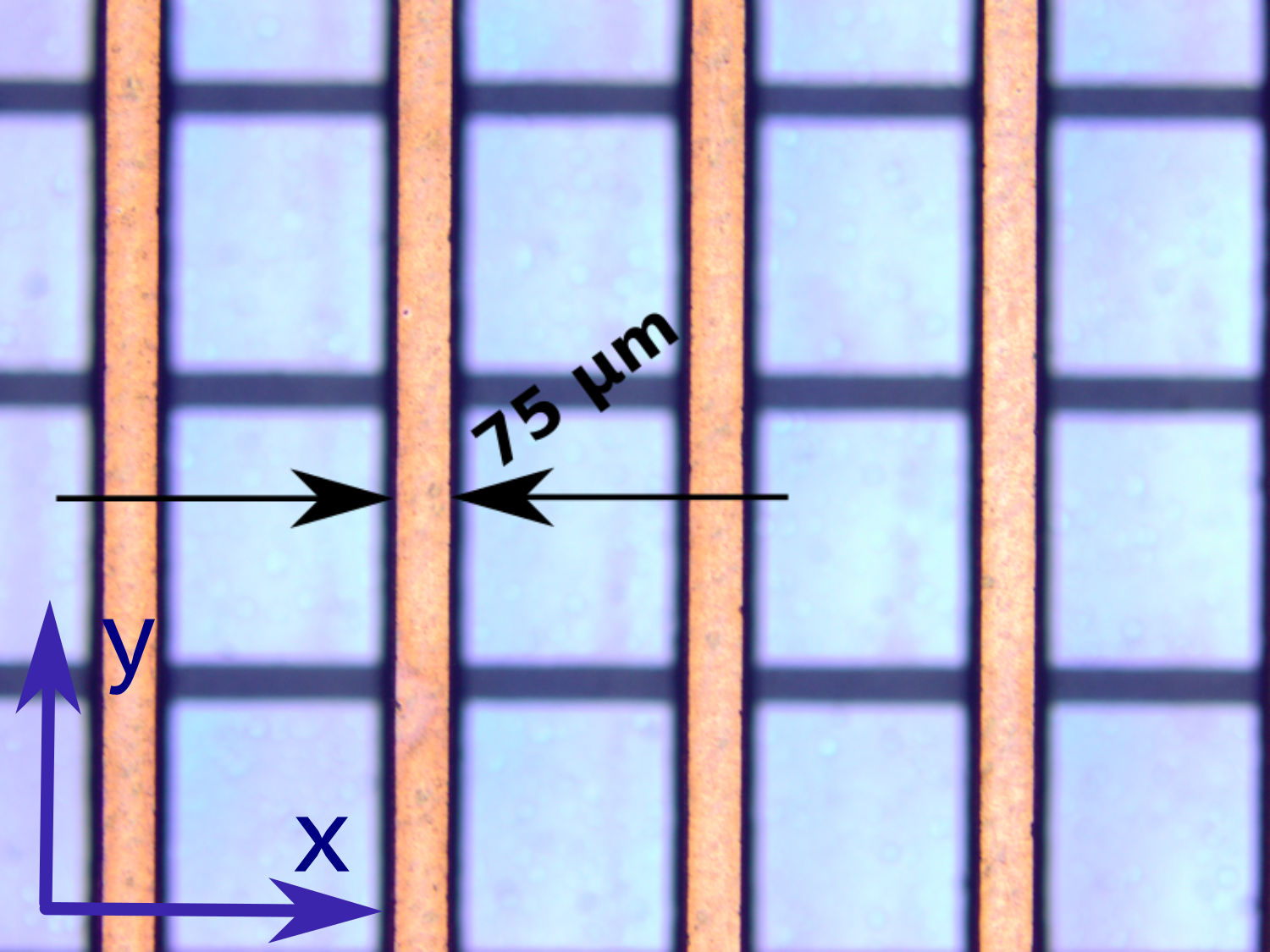} ~~~~~~~
	\includegraphics[height=4.5cm,keepaspectratio]{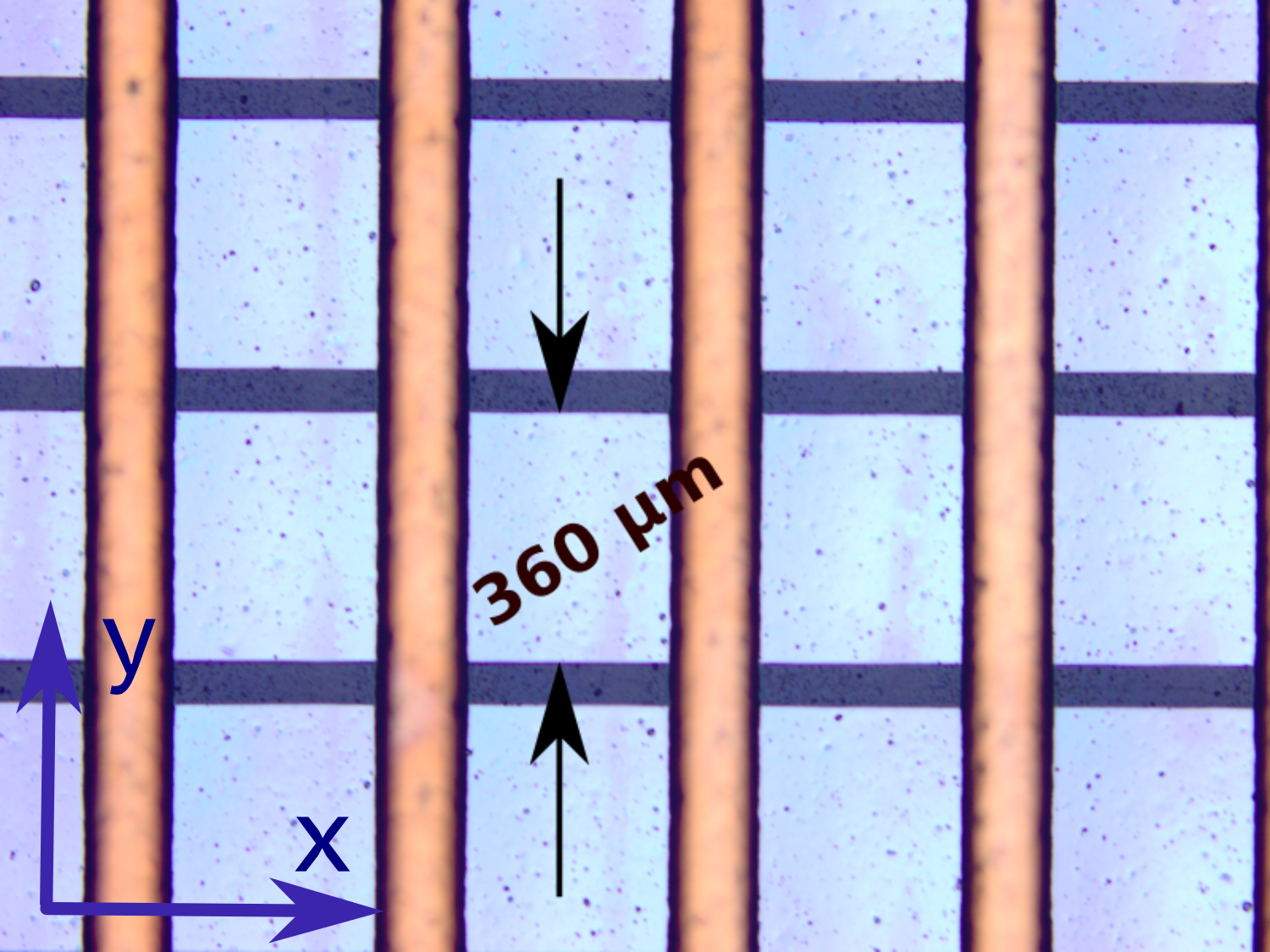}
	\caption{Microscopic picture of GEM readout with top layer in focus having 75 $\mu m$ strip width running along y-axis (left) and bottom layer in focus having 360 $\mu m$ strip width running along x-axis (right).}
	\label{GEM_Micro}
\end{figure}

The readout board has 256 strips running in both x and y planes with a pitch of 400~$\mu$m. The microscopic picture of the readout is as shown in figure~\ref{GEM_Micro}. These readout strips are bunched together and are connected to four 130 pin Panasonic connectors (the first and last pins are ground)  mounted on the readout PCB. For data taking, APV25 front-end boards~\cite{French:2001xb} have been used. Their function is to collect charge signals from the readout strips through Panasonic connectors.  The APV25 ASIC contains  pre-amplifier and shaper for each individual channels. The amplified analog signals are then transferred to a Scalable Readout System (SRS)~\cite{Martoiu:2011zja}, developed by the RD51 collaboration~\cite{Alfonsi:2008zz}, through an HDMI cable. The SRS has two main components: Analog to Digital Converter (ADC) and Front-End Concentrator (FEC)~\cite{toledo2011front}. The ADC converts the analog signals received from APVs into digital form, while the FEC handles all the communication, data processing and data transfer processes. 

The experiments have been performed with an Ar-CO$_2$ gas mixture having a volumetric ratio of 80-20, respectively. An $^{55}$Fe source with source strength of 8 $mCi$ has been used for data acquisition. 
The voltage and field configurations were as shown in table~\ref{table1} with single  GEM detector operating at foil voltage of 480~$V$ and double GEM at 380-380 and 380-400~$V$. The environmental parameters were closely monitored during the detector operation, and temperature and pressure have been maintained at 25~$\pm$~1\textdegree~C and 1005~$\pm$~6~hPa, respectively.



\begin{table*}[htbp]
\begin{center}
\caption{\bf Voltage/Field configurations} 
\scalebox{1.0}{
\begin{tabular}{ | c | c | c | c |  c | c |}
\hline
{\bf Description} & {\bf E$_{D}$ }& {\bf E$_{T}$ }&{\bf E$_{I}$ }&{\bf  $\Delta$V$_{GEM1}$ }&{\bf   $\Delta$V$_{GEM2}$}  \\
{ }& {\bf (kV/cm) }&{\bf  (kV/cm)} & {\bf (kV/cm) }&{\bf  (Volts) }& {\bf  (Volts)} \\
\hline
S-GEM & 2.76 &  & 3.33  & 480 &  \\
D-GEM$_1$ & 2.0 & 2.5 & 3.0 & 380 & 380 \\
D-GEM$_2$ & 2.0 & 2.5 & 3.0 & 380 & 400 \\
\hline
\end{tabular}}
 \label{table1} 
\\ ~\\Abbreviations for single and double GEM detectors are S, D-GEM respectively. 
\end{center}
\end{table*}

\section{Measurements}
\label{S3}
\subsection{Charge spread}
\label{S3a}
\begin{figure}[ht]
	\centering
	\includegraphics[width=0.485\linewidth,keepaspectratio]{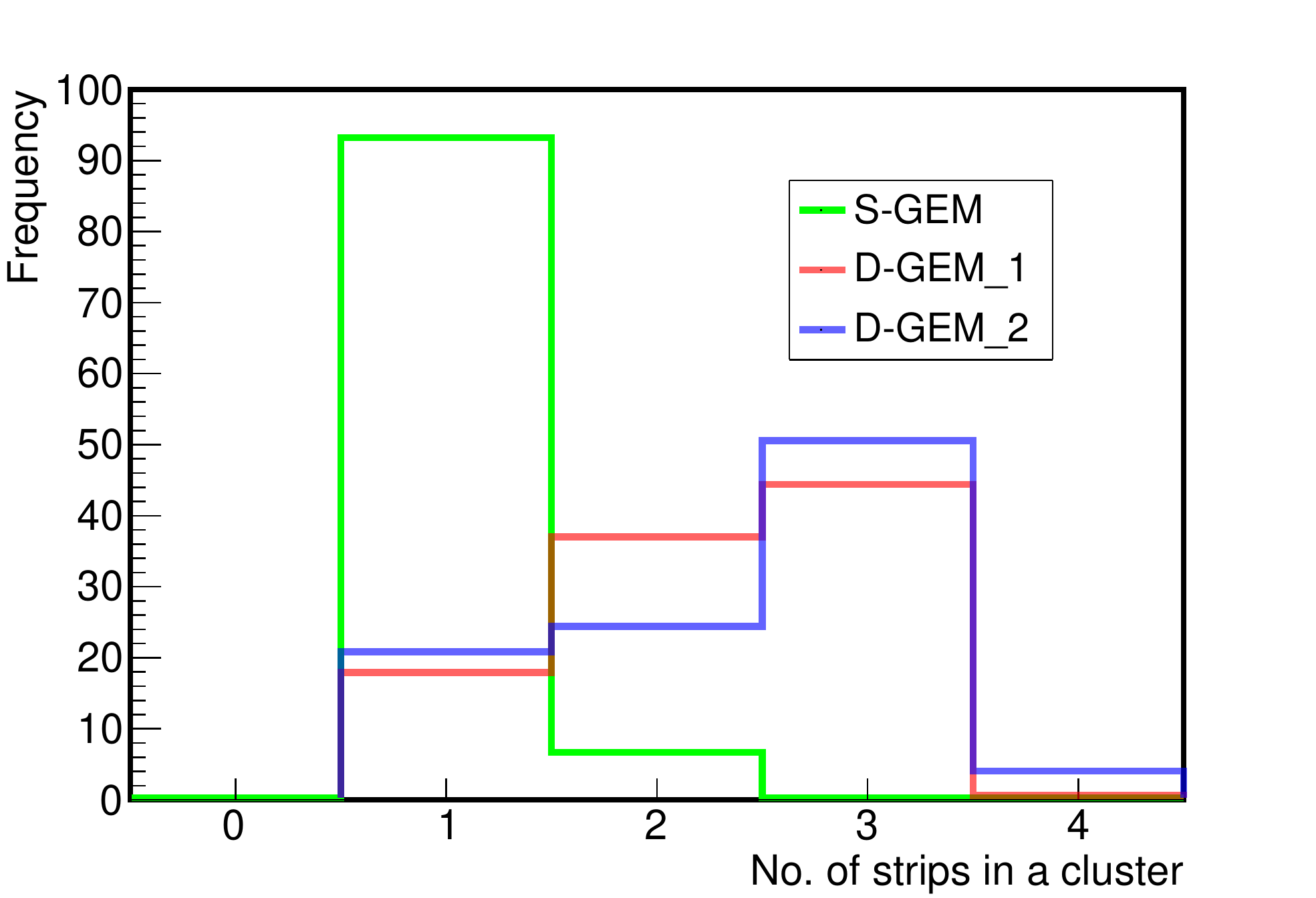}~
	\includegraphics[width=0.485\linewidth,keepaspectratio]{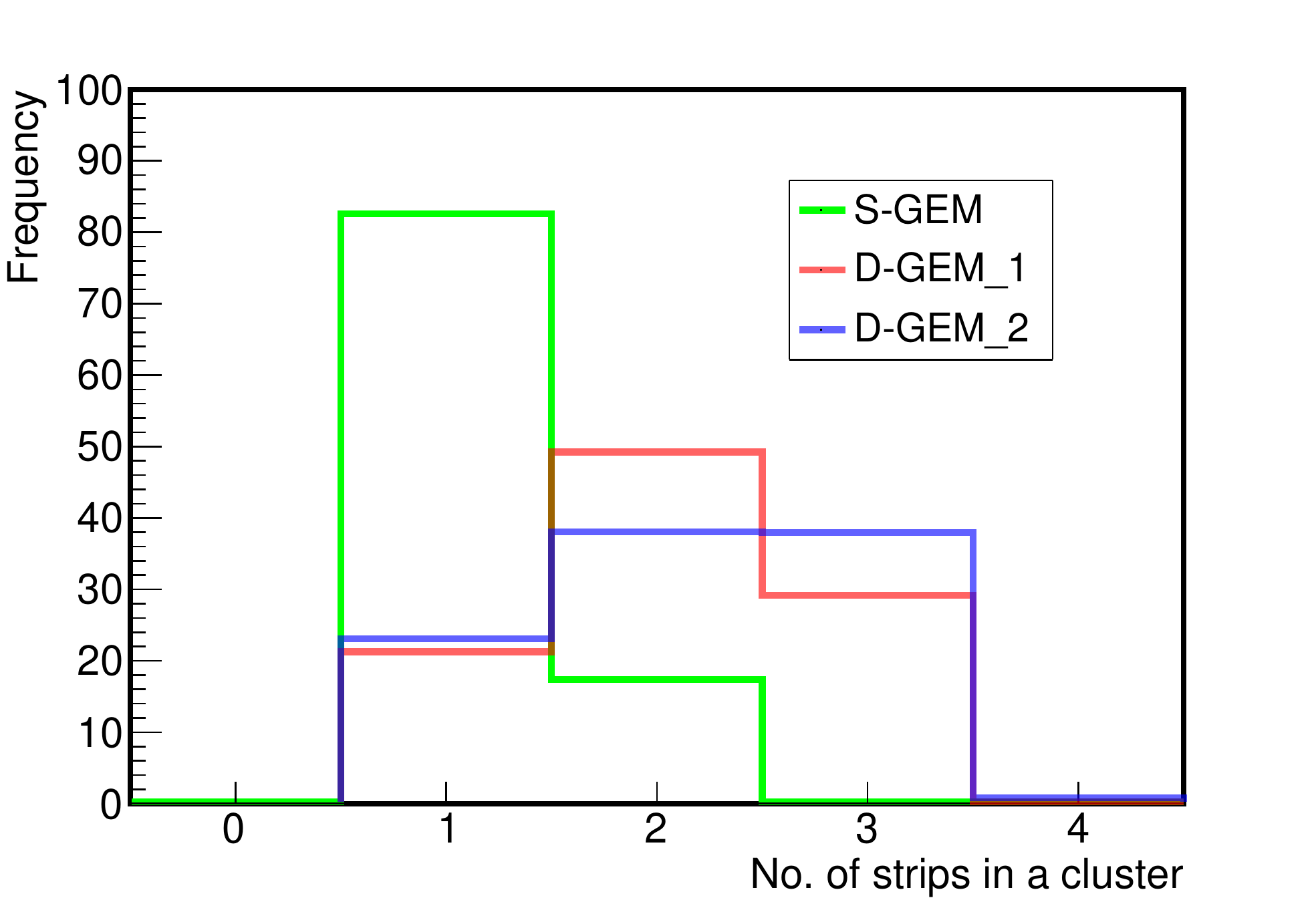}
	\caption{Strip multiplicity obtained experimentally for various GEM configurations with data from the x-sense plane (left) and y-sense plane (right).}
	\label{XY_Strip}
\end{figure}

After multiplication in the lower GEM foil (GEM$_1$), the electrons move towards the readout electrodes. The electrons created during a single x-ray event is termed as a cluster~\cite{Sauli:2016eeu} and the strip multiplicity of a cluster is determined by the number of strips 
that acquire signal above a threshold, for that cluster~\cite{abbaneo2014performance}. The spread of the cluster depends upon various parameters like the number of GEM foils, gas properties size of drift, transfer and induction gaps, electric field configurations and gain values~\cite{roy2021charge, abbaneo2014performance, maerschalk2016study}.  The charge spread information is derived experimentally from the number of strips hit in a single cluster event. The normalized frequency distribution of the number of strip hits per cluster (multiplicity) for configurations mentioned in table~\ref{table1} have been shown in figure~\ref{XY_Strip}. The strip multiplicity also depends upon the width of readout strips and its pitch i.e., 400~$\mu m$ in our case. 
The strip multiplicity for most of the events in a single GEM is one and is higher for double GEM configurations. This clearly shows that the spread in double GEM configuration is higher as compared to single GEM. Similar results have been observed by numerical simulations presented in~\cite{roy2021charge}. 
The plot also demonstrates that as the voltage is increased in one of the GEM foil from configuration D-GEM$_1$ to D-GEM$_2$, there is an increase in strip multiplicity.
Similar trends are also observed in~\cite{abbaneo2014performance} for triple GEM. 

\subsection{Position resolution} 
\label{S3b}
In a conventional method, the position resolution is determined by placing the test detector in parallel with several other position-sensitive detectors. The setup is used to measure the hit points of charged particles from a beamline or a cosmic ray. These hit points are used to determine the track of the charged particle using track-fitting algorithms. The track is then either interpolated~\cite{pellecchia2023performance} or extrapolated~\cite{zhang2019measurement} to determine the hit position of the track on the detector under observation. The position resolution is obtained by using residual analysis on the real hit and the hit obtained from the track. The detectors used for track reconstruction are either reference detectors with known position resolution~\cite{simon2009beam, barvich2002construction} or similar copies of the test detector~\cite{aulchenko2009triple, ketzer2004performance} under observation to obtain convoluted position resolution of all the detectors. The use of multiple detectors increases the complexity of the measurment.
To counter this problem,     
another commonly used method is  to use a fine collimated slit to radiate the detector using a source~\cite{zhou2009study}. 
In order to overcome problems related to accurate alingnment of slit and influence of slit width on position resolution measurment, many variations on the same approach have been proposed. For example, to reduce
 the mechenical complexity a variation of slit width or introduction of sharp edge instead of slit has been tried~\cite{greer2000evaluation, samei1998method}.

\begin{figure}[ht]
	\centering
	\includegraphics[width=0.56\linewidth,keepaspectratio]{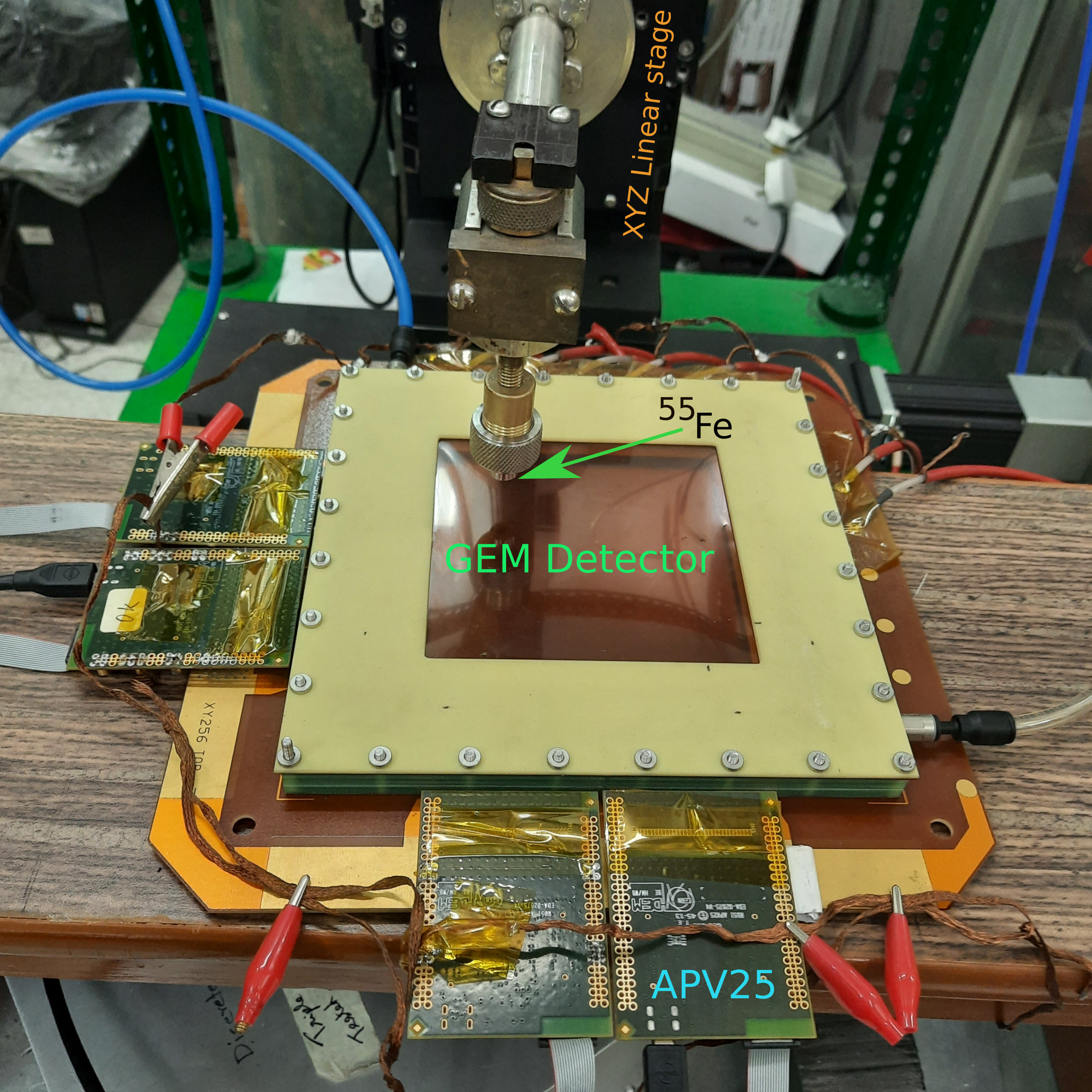} ~~
	\includegraphics[width=0.4\linewidth,keepaspectratio]{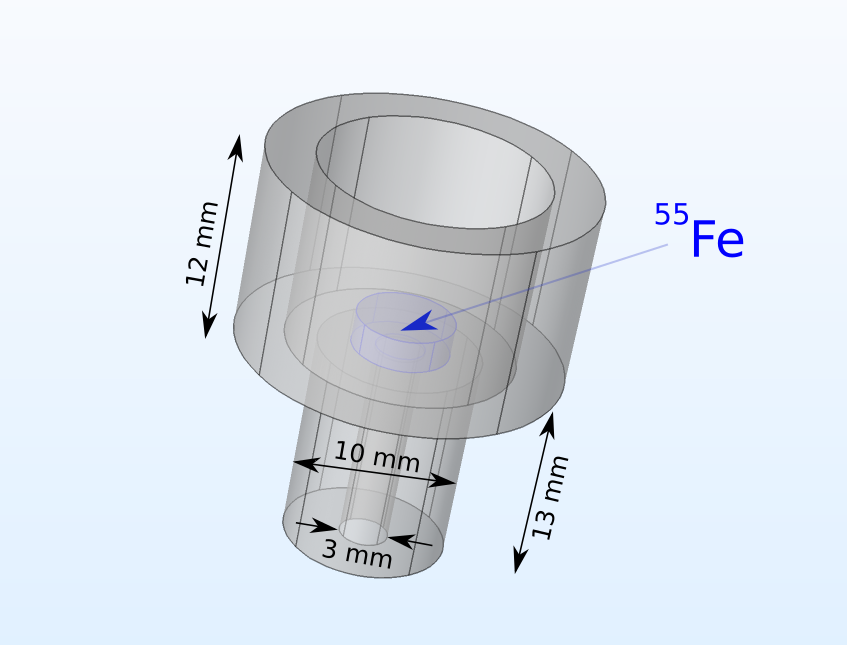}
	\caption{ Image of experimental setup for position resolution measurement (left) and the schematic of collimator along with the source (right).}
	\label{GEM_Post_setup}
\end{figure}

In the present work, to determine the position resolution,  an experiment has been performed by placing the detector on a test bench, where the $^{55}$Fe x-ray source has been mounted on AEROTECH PRO165 triple-axis XYZ Linear Stage~\cite{Aerotech}. The Linear Stage can move in x, y and z-directions with 0.5~$\mu m$ resolution. The source is placed in a collimator made of stainless steel  with collimation length 13~$mm$, having an inner and outer diameter of 3 and 10~$mm$, respectively.  This ensures that the source is well collimated with most of the radiation beam falling perpendicular to the detector plane. The experimental setup and a schematic diagram of collimator are shown in figure~\ref{GEM_Post_setup}.  

\begin{figure}[ht]
	\centering
	\includegraphics[width=0.47\linewidth,keepaspectratio]{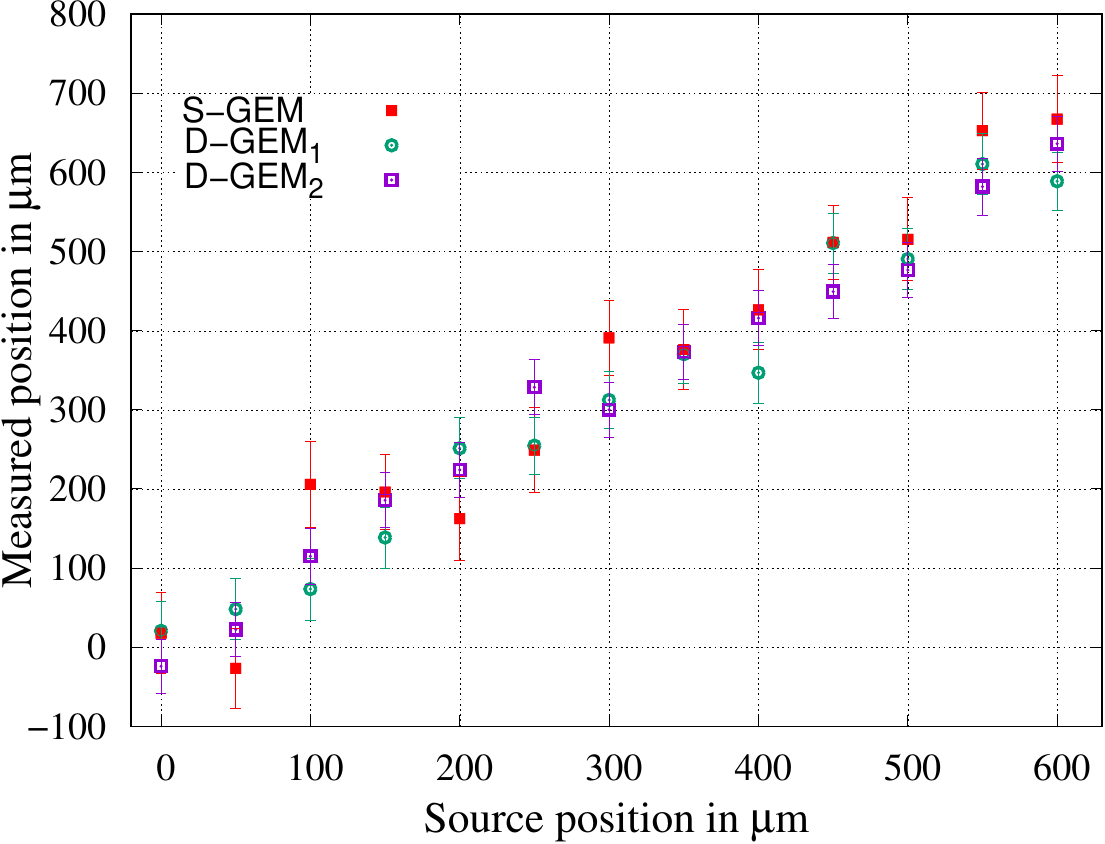}~~~
	\includegraphics[width=0.47\linewidth,keepaspectratio]{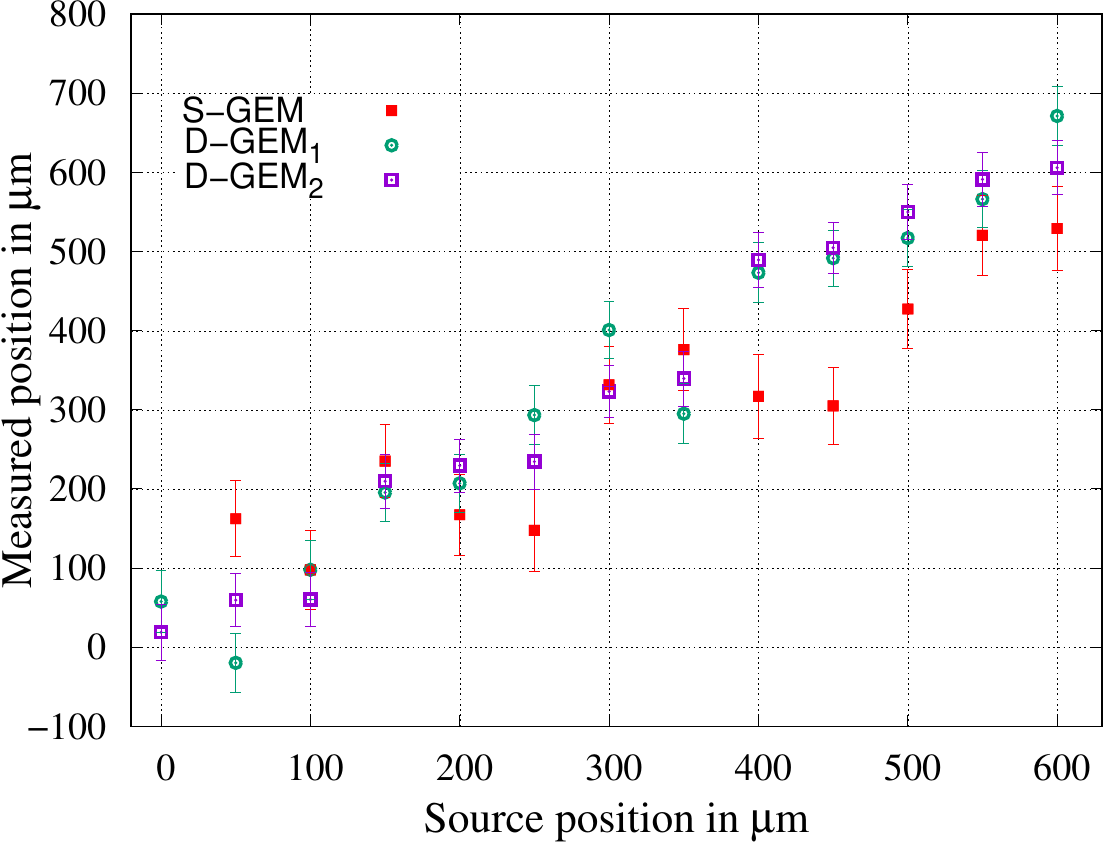}
	\caption{Detector response with source movement in diagonal  direction in 50 $\mu m$ steps. Measurments obtained from x-sense plane (left) and
	 y-sense plane (right).}
	\label{XY_pos}
\end{figure}

The experiment has been performed by using the Linear Stage to move the source in a step of 50~$\mu m$ in both x and y direction, diagonally.   The data has been recorded after each step,  with the APV25 front-end boards. A BASIC code controls the movement of the Linear Stage, and its purpose is to ensure that the source is moved appropriately after the data has been recorded by the DAQ system at a specific spot. The position information of each cluster is determined  by evaluating the central value of the charge spread.  For this,  the Centre of Gravity (CoG) method ( i.e. a generalization of Anger logic~\cite{anger1964scintillation}) has been used where charge sharing information is utilized to give weights to the strips. It has to be noted that the CoG method works effectively for multiplicity levels of three and above and fails for single strip events.
The source position information from the readout is obtained by fitting a Gaussian on all the event points recorded at one position. Figure~\ref{XY_pos} shows the movement of source position recorded by the GEM detector with respect to the actual movement of the source by Linear Stage.
The spatial resolution at one sigma level (standard deviation) for all the three configurations are mentioned in table~\ref{table2} i.e., the standard deviation of the spread obtained by plotting the difference between the true value and the experimental value of position from the repeated measurements.


\begin{table}[htbp]
\begin{center}
 \caption{\bf Position resolution of GEM detector} 
\scalebox{1.0}{
\begin{tabular}{ |c|c|c|}
\hline 
{\bf Detector}&  {\bf X-Plane Position } & {\bf Y-Plane Position}     \\
{\bf Configuration }&{\bf Resolution ($\mu m$)}& {\bf  Resolution ($\mu m$)  } \\
\hline
S-GEM& 66.47 (20.04)& 88.86 (26.79)\\
D-GEM$_1$& 57.41 (13.17)& 68.499 (15.71)\\
D-GEM$_2$ & 36.76 (8.43)& 54.56 (12.86)\\
\hline
\end{tabular}}
 \label{table2} 
 \\~
\\ *Error associated with the measurement are shown in brackets.
\end{center}
\end{table}

The simulation result using the Monte-Carlo method reported in~\cite{lan2013study} gives single GEM spatial resolution of $\sim$~58~$\mu m$ for standard GEM foil and Ar-CO$_2$ 80-20 gas mixture. In~\cite{wang2013practical} spatial resolution of double GEM is determined experimentally in a gas mixture of Ar-CO$_2$ 85-15 and is reported to be 56~$\pm$~15 $\mu m$ for a standard GEM foil with one-dimensional readout having a pitch of 200~$\mu m$. 

The method employed for position resolution measurment in this work offers a significant advantage over conventional techniques, as it necessitates only an equipment capable of precise movement. This novel approach does away with the need for cumbersome stacks of position-sensitive detectors or a tricky-to-manage slit, which can introduce a range of complexities such as the demanding precision of slit width and can cause diffraction effects~\cite{zhang2017spatial} that can lead to inaccuracies.

\subsection{Gain and energy resolution}
\label{S3c}
\begin{figure}[ht]
	\centering
	\includegraphics[width=0.485\linewidth,keepaspectratio]{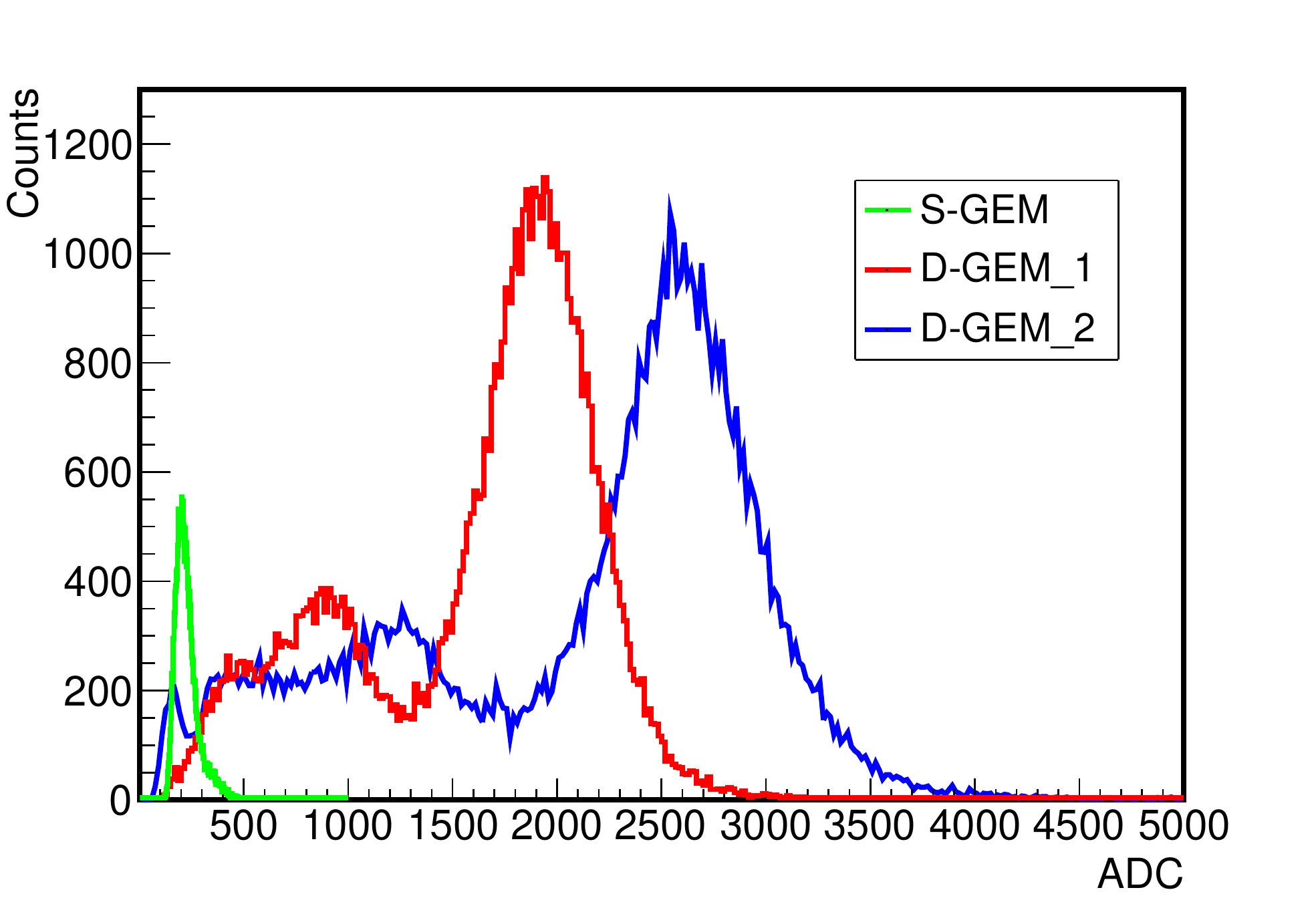}~
	\includegraphics[width=0.485\linewidth,keepaspectratio]{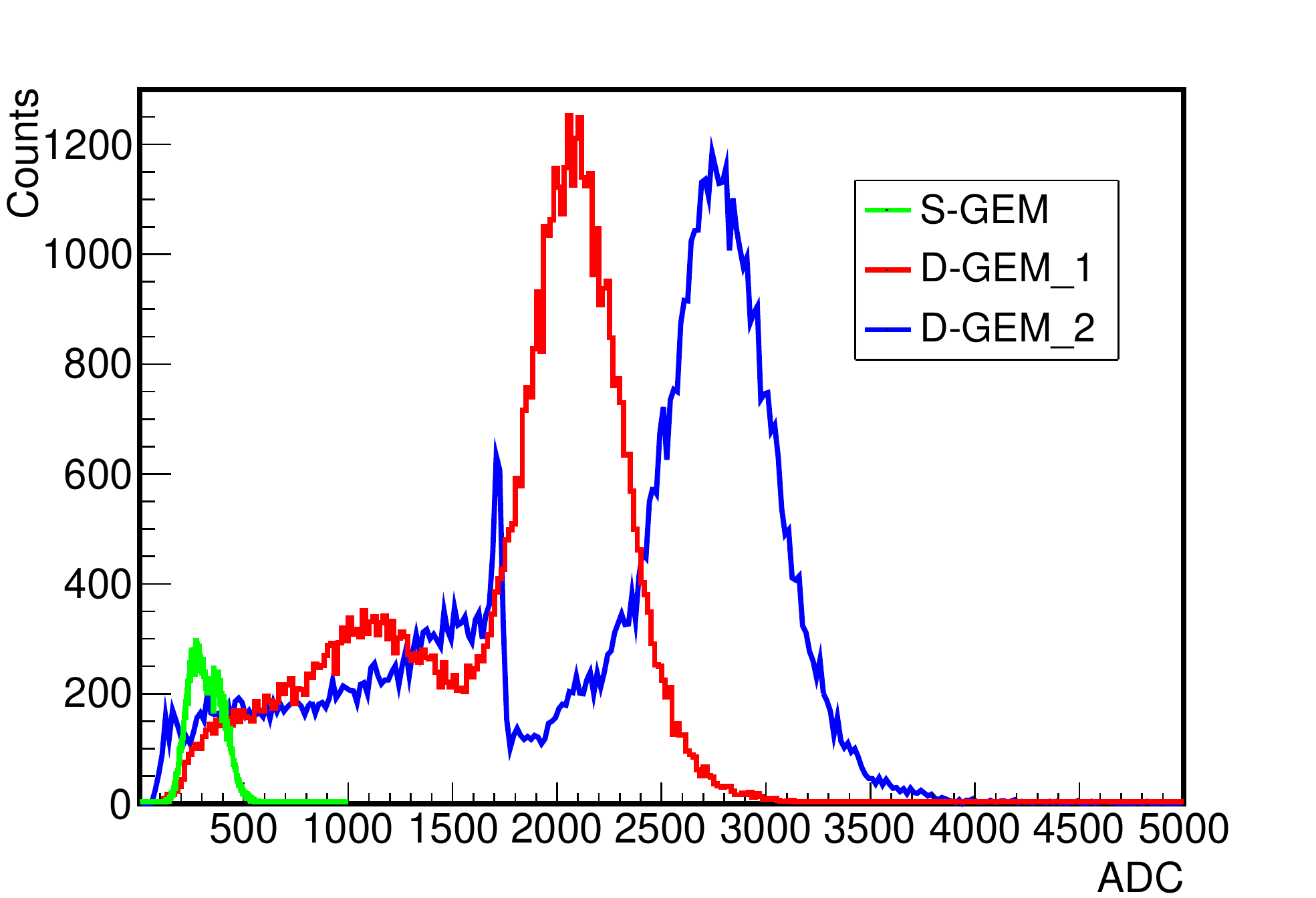}
	\caption{Energy spectra obtained from x-sense plane (left) and y-sense plane (right). }
	\label{XY_EnergySpec}
\end{figure}

The energy spectra have been obtained by plotting a histogram of the total charge collected in an event cluster, i.e., the sum of charges from individual strips in a cluster. The energy spectra for all the configurations of our study have been shown in figure~\ref{XY_EnergySpec}.
 The gain, as expected, is much lower for single GEM than that of double GEM. However, the voltages applied to the single GEM foil  (480~V) are much higher than the voltages applied in double GEM foils~\ref{table1}. The  double GEM configuration increases the detector gain substantially  and has a much lower discharge probability due to the low working voltage of the GEM foils~\cite{Bachmann:2000az, rout2021numerical}. The energy spectra obtained from the y-sense plane of D-GEM$_2$ have a spike at $\sim$1700 ADC value which is due to the gain saturation of the strip. This issue can be resolved either by reducing the foil voltage or by reducing the electronics noise (pedestal) in the APV25 board.  For energy resolution measurement, the 5.9~$keV$ $^{55}$Fe photo-peak is fitted with the Gaussian function and the results are mentioned in table~\ref{table3}.
\begin{table}[htbp]
\begin{center}
\caption{\label{table3} \bf Gain value measured from x-sense plane}
\scalebox{1.}{
\begin{tabular}{ |c|c|c|c|}
\hline 
{ \bf  Detector}& { \bf   Mean (G)} & { \bf Sigma ($\sigma_G$) }& { \bf  $\sigma_G/G~\%$}  \\
\hline
S-GEM& 210.75 (0.29)& 37.02 (0.26)&17.57\\
D-GEM$_1$& 1915.72 (1.83) & 263.64 (2.347)&13.76\\
D-GEM$_2$ & 2598.93 (2.33) & 319.40 (3.10)&12.29\\
\hline
\end{tabular}}
 \\~
\\ *Error associated with the measurement are shown in brackets.
\end{center}
\end{table}



\subsection{Gain uniformity}
\label{S3d}
\begin{figure}[ht]
	\centering
	\includegraphics[height=7cm,keepaspectratio]{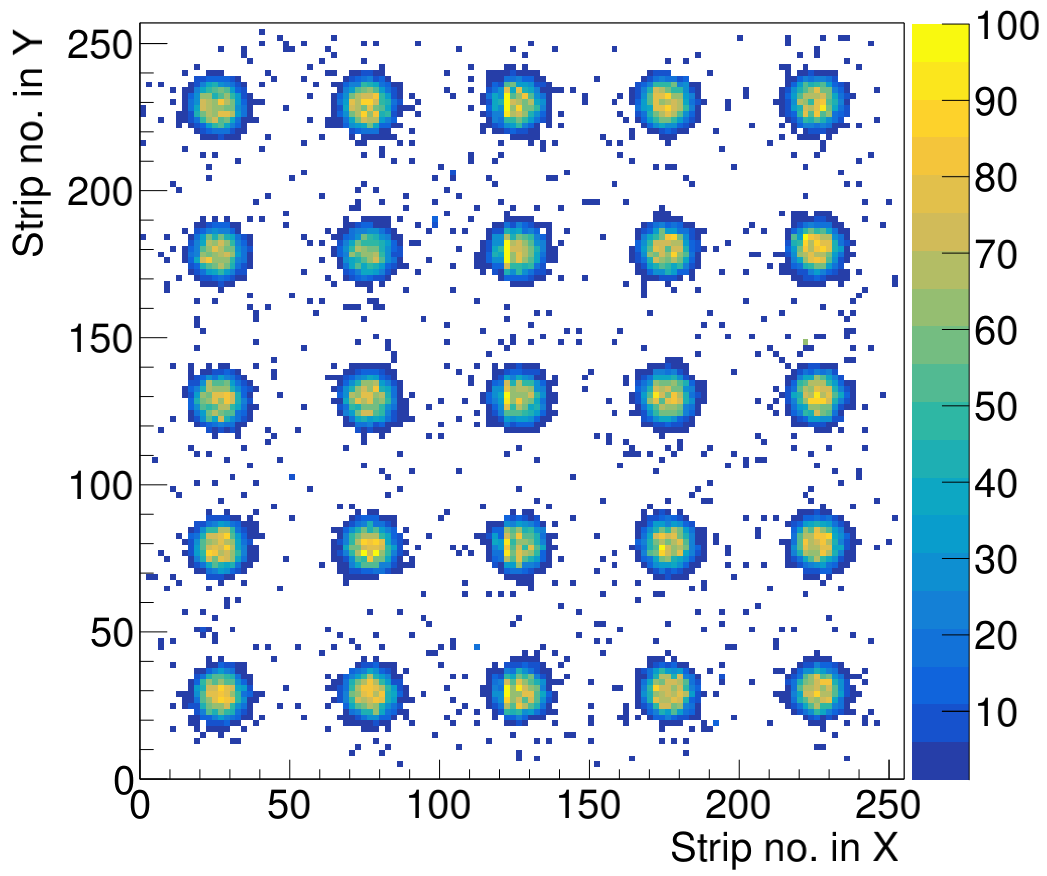}
	\caption{Hit map obtained by placing the source at the center of each  sub-area to measure gain
	    The color represents the total number of hit events.}
	\label{HitPlot}
\end{figure}

To measure the gain uniformity of the detector, it is virtually divided into 25 equal sub-areas, each consisting of 5 rows and 5 columns. The source is then placed at the center of each sub-area to measure the detector gain in that region. The resulting hit map is shown in Figure \ref{HitPlot}.

\begin{figure}[ht]
	\centering
	\includegraphics[width=0.48\linewidth,keepaspectratio]{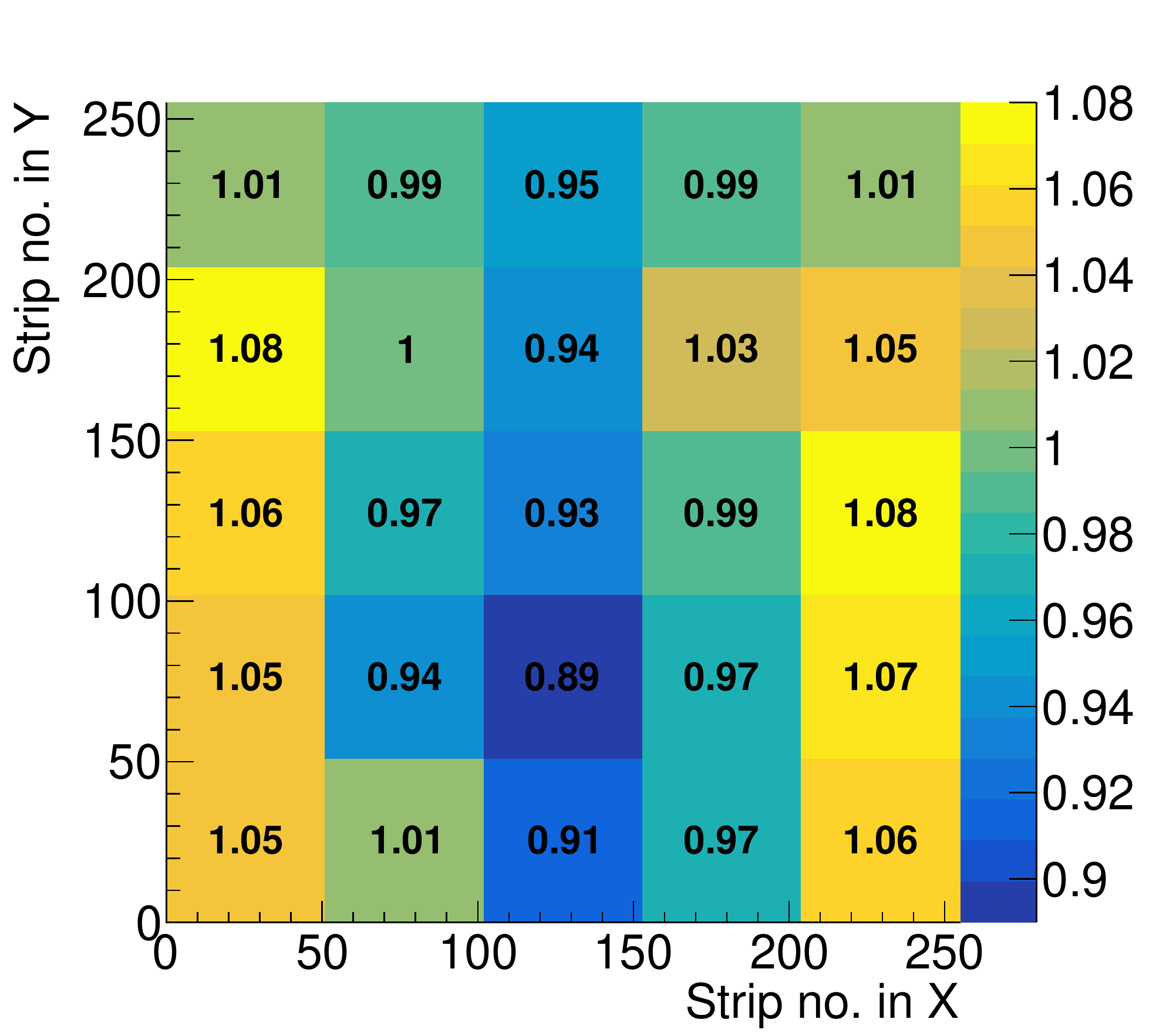} ~~~
	\includegraphics[width=0.477\linewidth,keepaspectratio]{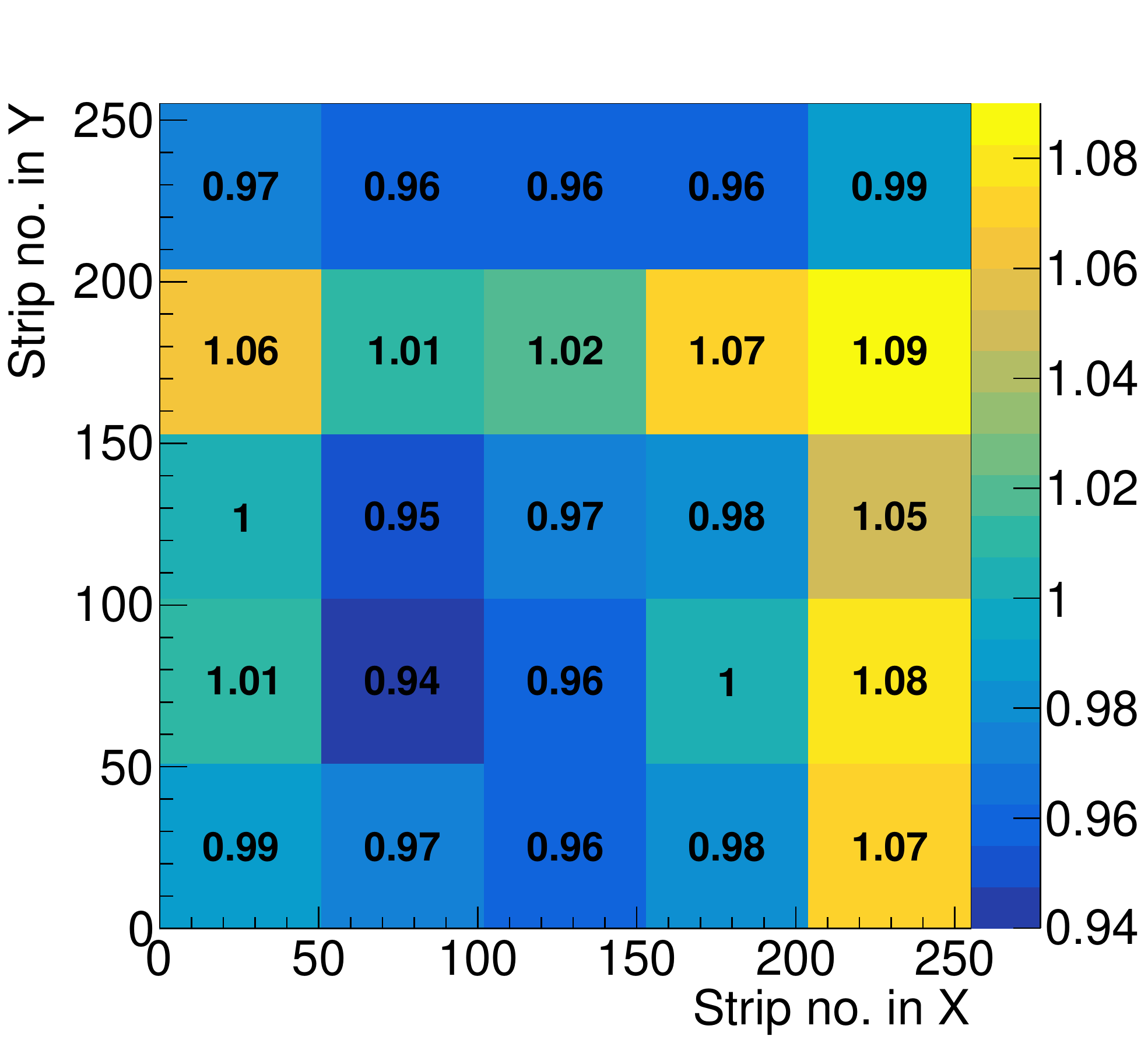}
	\caption{Gain map obtained from x-sense plane (left) and y-sense plane (right).}
	\label{GainMap}
\end{figure}

\begin{figure}[ht]
	\centering
	\includegraphics[height=6.5cm,keepaspectratio]{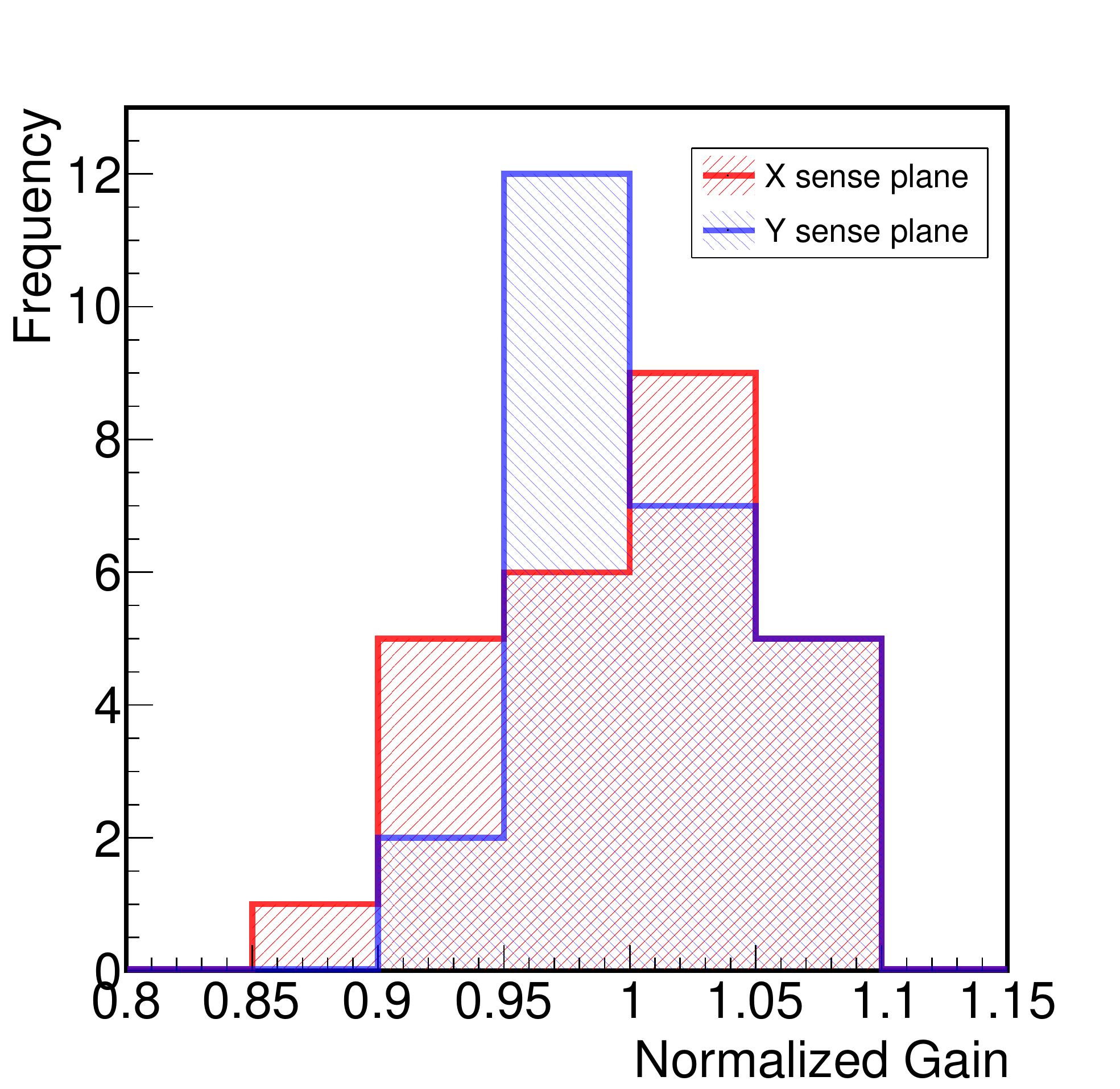}
	\caption{Gain map distribution.}
	\label{GainFreq}
\end{figure}

The spectra for all the sub-areas have been analyzed to obtain the gain uniformity map, as shown in Figure \ref{GainMap}. The gain is found to be reasonably constant, with a variation from a maximum of 1.08 to a minimum value of 0.90. Figure \ref{GainFreq} shows the histogram of normalized gain, and the gain values of most regions are close to the mean value of 1.0. The results are consistent with double GEM results presented in \cite{yu2003study} with similar detector size and gas mixture. They are also in agreement with triple GEM results reported in \cite{bazylev2018study, ketzer2002triple, Chatterjee:2018dxm, Patra:2017gaz}.


\section{Numerical simulation}
\label{S4}
\begin{figure}[ht]
	\centering
	\includegraphics[width=0.485\linewidth,keepaspectratio]{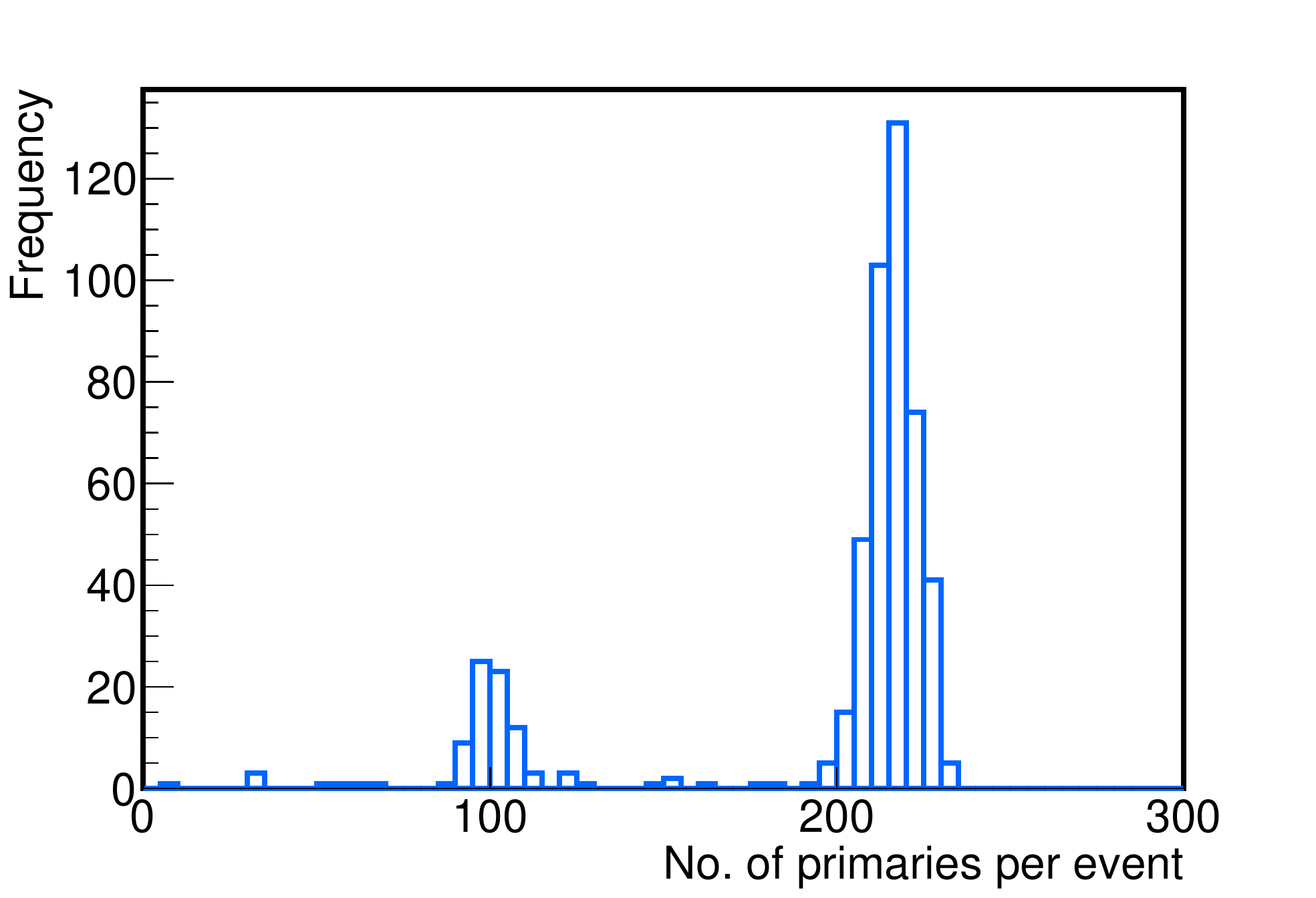}~~
	\includegraphics[width=0.485\linewidth,keepaspectratio]{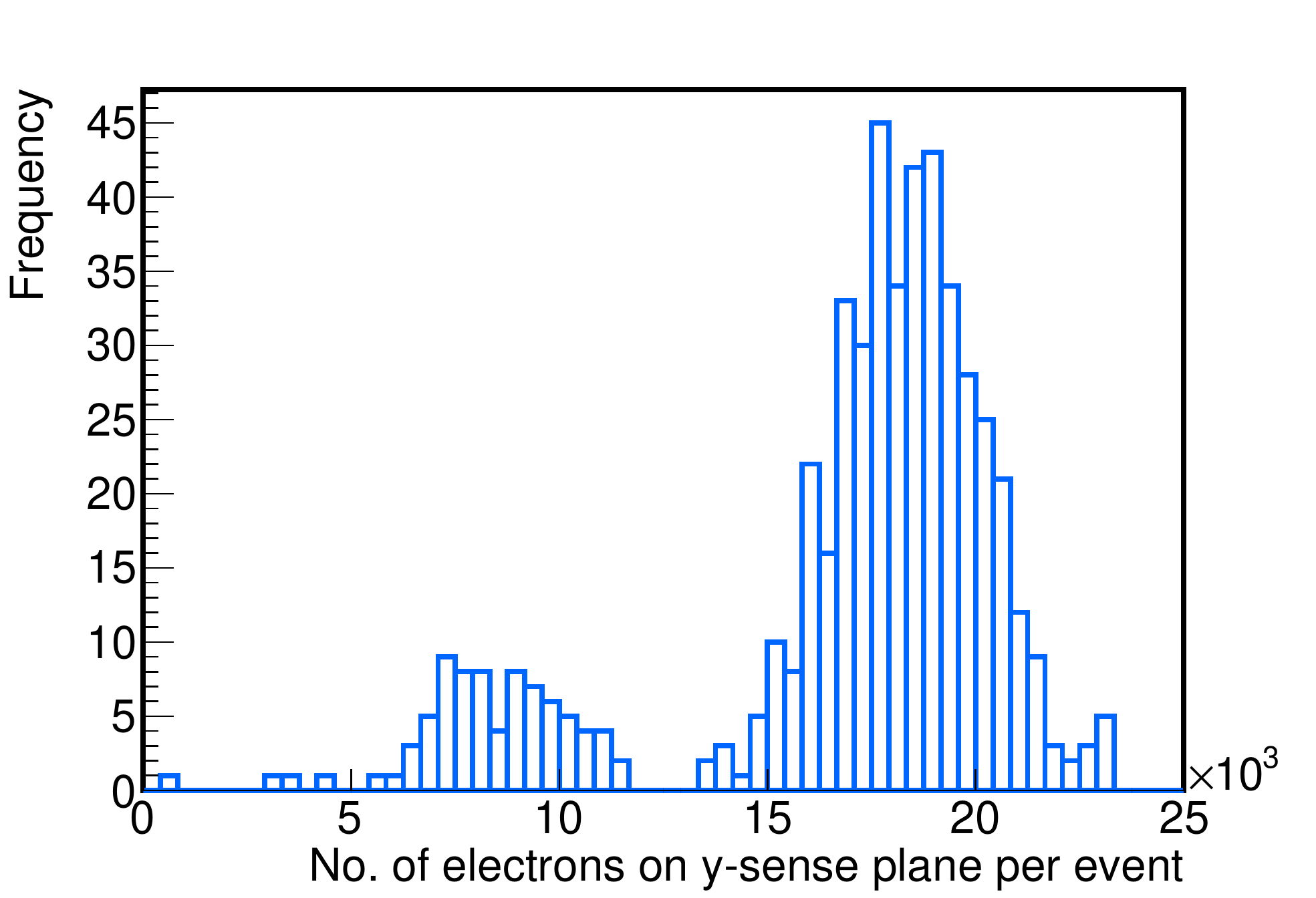}
	\caption{Number of primaries per  $^{55}$Fe x-ray event generated by HEED for single GEM configuration mentioned in table~\ref{table1} (left) and the corresponding number of electrons collected from the y-sense plane (right).}
	\label{Ch5:PrimGain}
\end{figure}

To verify the strip multiplicity results obtained from the experiment section~\ref{S3a}, a simulation work has been conducted using the GARFIELD simulation framework. The numerical model has been made identical to the experimental setup by providing a similar geometry, gas mixture, temperature, pressure, voltage configuration, etc.
Primary ionization information and transport properties for $^{55}$Fe x-ray tracks have been obtained using HEED~\cite{Smirnov:2005yi} and MAGBOLTZ~\cite{Biagi:1999nwa}, respectively. The electric field configuration has been computed using neBEM. The $^{55}$Fe tracks have been used for generating the primaries. Figure~\ref{Ch5:PrimGain} (left panel) shows the frequency distribution of number of primaries generated during $^{55}$Fe events. The primary electrons are then allowed to transport from the drift volume to the anode plane after multiplication in the GEM foil(s) and are collected by the x and y-sense planes.
The frequency distribution of number of electrons collected on the y-sense plane per event is shown in figure~\ref{Ch5:PrimGain} (right panel).
It should be noted that in order to reduce computational time the Penning effect was not considered in these calculations.

\begin{figure}[ht]
	\centering
	\includegraphics[width=0.485\linewidth,keepaspectratio]{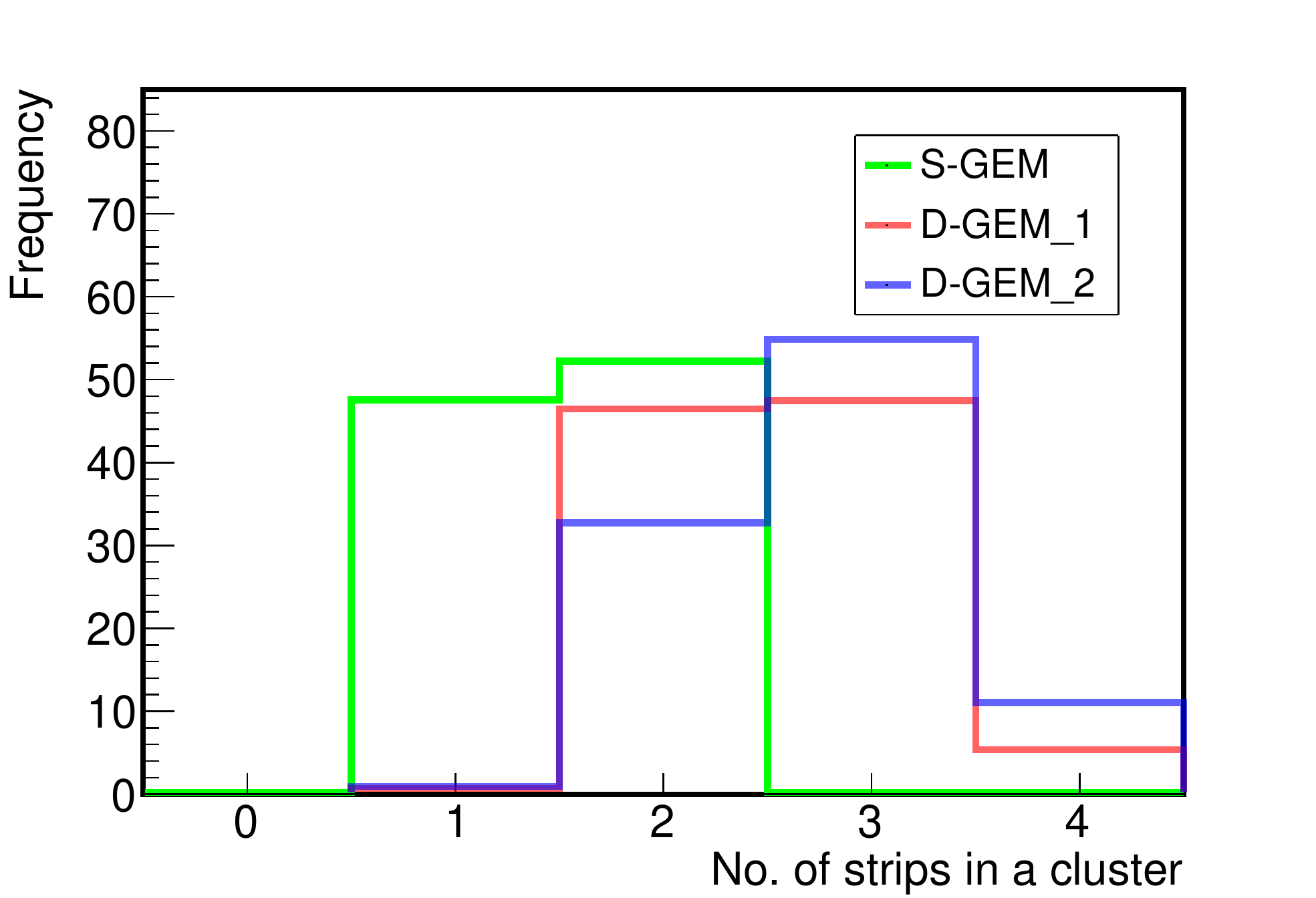}~~
	\includegraphics[width=0.485\linewidth,keepaspectratio]{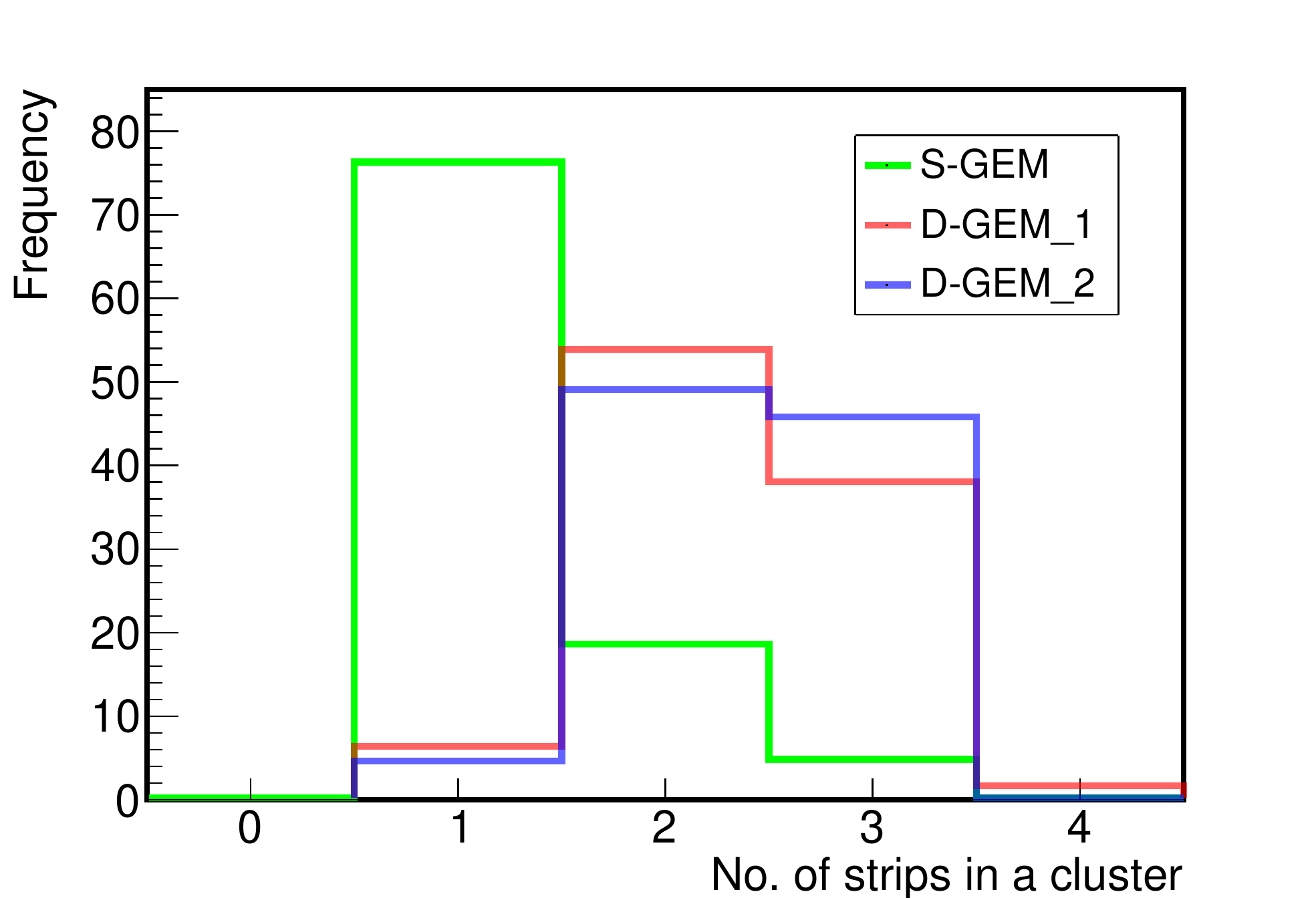}
	\caption{Strip multiplicity obtained from simulation for single and double GEM configurations from x-sense plane (left) and y-sense plane (right).}
	\label{Ch5:SimStrip}
\end{figure}

The collection of the electrons reaching the readout has been estimated based on a readout strip configuration as in figure~\ref{GEM_Micro}. The strip is considered to have a hit if it has at least 10$\%$ of the total number of electrons that are detected in an event. The results of the simulation work are shown in figure~\ref{Ch5:SimStrip}. A similar trend has been observed as seen from the experimental measurements (figure~\ref{XY_Strip}) which is an increase in strip multiplicity with gain and with increase in GEM foils. However, due to differences like charging-up, Penning effect and induction of charge in the readout strips that are not considered in simulation, the results are not an exact match. 

\section{Summary}
\label{S6}
A comprehensive and meticulous investigation into the charge spread, position resolution, energy resolution, and gain uniformity of single and double GEM configurations has been conducted through a series of experiments. Employing state-of-the-art tools such as an APV25 front-end chip, an SRS-based DAQ system, and an $^{55}$Fe source, the study has yielded insightful results. The charge spread information obtained from the strip multiplicity shows a similar trend as  obtained from the Garfield simulation for a similar configuration, validating our results.
As the number of GEM foils and their corresponding voltage values increase, the size of the charge cluster also escalates, leading to a corresponding increase in strip multiplicity.

In an effort to measure the spatial resolution, a high-precision positioning instrument was utilized for source movement, which turned out to be highly successful. The double GEM configuration, in particular, demonstrated exceptional ability in resolving positions with sigma values up to 36.76~$\mu m$ and 54.56~$\mu m$ in x and y directions, respectively. While the spatial resolution for a single GEM is also commendable, estimated to be less than 100 $\mu m$, the double GEM detectors offer several advantages, including low discharge probability due to low GEM foil voltage and higher efficiency due to higher gain, which are particularly crucial for detecting weakly interacting particles like muons.

Another crucial aspect examined in this study is the gain uniformity of the GEM detector. The findings are  in good agreement with previous works on double GEM~\cite{yu2003study} and are in consistent with prior results of triple GEM, providing further validation for the study's results. The gain of the detector serves as an indicator of its sensitivity throughout the active region, and the results show that it has been consistent, with only minor deviations from the mean value.

\acknowledgments
The authors would like to thank Mr. Shaibal Saha for his technical help in assembling the GEM detectors. We would also like to acknowledge Dr.~Jaydeep Datta, Mr.~Pralay Das, Dr.~Sridhar Tripathy and Dr.~Prasant K. Rout for their valuable suggestions and advice in experimental work and data analysis. We would also like to thank Prof. Sudeb Bhattacharya and Prof. Chinmay Basu for their support. This work has partly been performed in the framework of the RD51 Collaboration and we would like to acknowledge the members of the RD51 Collaboration for their help and suggestions.

\bibliographystyle{unsrt}
\bibliography{Main}

\end{document}